# Carbon Irradiated SI-GaAs for Photoconductive THz Detection

*Abhishek Singh\*, Sanjoy Pal, Harshad Surdi, S. S. Prabhu, Mathimalar S., Vandana Nanal, R. G. Pillay and G.H. Döhler*

Abhishek Singh, Sanjoy Pal, Harshad Surdi, S. S. Prabhu, Mathimalar S., Vandana Nanal and R. G. Pillay
*TATA Institute of Fundamental Research, Mumbai, 400005, India*
G.H. Döhler
*Max Planck Institute for the Science of Light, Erlangen, Germany*
E-mail: (asingh@tifr.res.in)

**Abstract:** We report here a photoconductive material for THz generation and detection with sub-picosecond carrier lifetime made by $C^{12}$ (Carbon) irradiation on commercially available semi-insulating (SI) GaAs. We are able to reduce the carrier lifetime of SI-GaAs down to sub-picosecond by irradiating it with various irradiation dosages of Carbon ($C^{12}$) ions. With an increase of the irradiation dose from $\sim 10^{12}$ /cm$^2$ to $\sim 10^{15}$ /cm$^2$ the carrier lifetime of SI-GaAs monotonously decreases to 0.55 picosecond, resulting in strongly improved THz pulse detection compared with normal SI-GaAs.

Electromagnetic radiation having frequencies in Tera-hertz (THz) range (1THz = $10^{12}$ Hz) have potential applications in security imaging, bio-sensing, chemical identification, material characterization etc. [1-4]. These materials interact with the passing THz radiation, imprinting specific spectral features to it. Due to its long wavelength features, the spatial resolution of THz radiation is less compared to visible-IR waves, but the information gained from it is enormous. For a faithful detection of the THz pulses, it is necessary that the detectors (1) are not adding extra detector related features and (2) are sensitive at the same time. This requires that the detector material should not exhibit any absorption lines in the THz spectrum and the antenna design should be free of any sharp resonance features in the detection frequency range.

Frequencies in the range of 1 THz are too high for conventional electronic devices. At the same time the photon energy of 4.2 meV corresponding to 1 THz is too low for conventional devices. Therefore, both approaches are facing enormous challenges for generating or detecting the THz radiation, in particular at room temperature. There are only a few techniques for generation and detection of THz radiation. Among those, the photoconductive technique is the most efficient and popular one [5, 6]. But the lack of suitable semiconductor materials represents a major restriction. In spite of being a direct bandgap material with high absorption in the near infrared (NIR) range, *standard* GaAs does not represent a good choice for photoconductive devices at THz frequencies because of its long lifetimes for photogenerated carriers. *Low temperature grown* GaAs (LT-GaAs), however, has turned out to be the most suitable material for fabrication of THz photoconductive sources and detectors because of its subpicosecond carrier lifetimes [7]. Improving the efficiency and reducing the manufacturing cost of these sources-detectors is one major goal of current research in the area. Growth of LT-GaAs is itself a difficult and expensive technology. Semi-insulating GaAs (SI-GaAs) is an economical alternative to LT-GaAs but it has the disadvantage of a much higher carrier lifetime (~ 100 ps) which makes it unsuitable for faithful THz pulse detection. In an earlier paper [8], we have reported on the advantages of carbon irradiated SI-GaAs for photoconductive THz *sources* compared with those made from standard SI-GaAs. Here, we present drastic improvement in the *detection* of THz pulses using these irradiated substrates. We observe a monotonous reduction of the carrier lifetime of SI-GaAs from ~70 ps down to ~0.55 ps by irradiation dosages of Carbon ($C^{12}$) ions ranging from ~$10^{12}$ /cm$^2$ to ~$10^{15}$ /cm$^2$. As a result the detection signal becomes increasingly faithful reproduction of the THz pulse. At the same time, the amplitude of the detection signal, the signal-to-noise ratio and the detection bandwidth increase drastically compared with normal SI-GaAs.

The irradiation was carried out using a 33.5 MeV beam of $^{12}$C from the Pelletron Linac Facility, Mumbai. The beam was passed through a ~10 μm thin gold foil which acted as an energy degrader and also generated the energy spread. Optimization of the gold foil thickness and the incident beam energy was done using the software SRIM [www.srim.org] to get a nearly uniform distribution of defects up to ~ 2 μm depth inside the SI-GaAs crystal. Two parallel metal electrodes separated by 25 μm were fabricated using standard photolithography technique on (un-annealed) irradiated and non-irradiated substrates of SI-GaAs. A schematic diagram is shown in fig.1.

In photoconductive detection of THz pulses, the near infra-red (NIR) and THz pulses are focused on the photo-conducting substrate with a variable relative delay time τ. The NIR pulses generate charge carriers, which move under the influence of the electric field caused by the THz pulse. The photocurrent is collected by the two electrodes. The measured photocurrent, *I(τ)* depends on the delay time between the NIR and the THz pulse. It is given by the integral of the product of the time dependent carrier density, *n(t-τ)*, and the time dependent electric field of the THz pulse *E(t)*, i.e.

$$I(\tau) \propto \int_{-\infty}^{\infty} E(t).n(t-\tau)dt \qquad (1)$$

As the NIR pulse is very short (<< 1 ps), there is a quasi-steplike increase in the charge carrier density *n* in the photoconductor. However, depending on the carrier lifetime of the material, *n(t-τ)* will decay either slowly (like in SI-GaAs) or rapidly (like in LT-GaAs). In the case of very rapid decay, the response function *n(t-τ)* assumes a delta-function-like form compared to the relatively slowly varying THz field *E(t)*. Hence, the photo current *I(τ)* measured in the case of a delay τ between arrival of the THz field and the probe pulse will be directly proportional to the electric field of the THz pulse, *E(τ)*, present at the photoconductor surface at the delayed carrier excitation by the NIR probe pulse. By changing the delay between arrival time of the NIR pulse relative to the THz pulse the whole THz pulse profile, *E(τ)*, is recorded. If the carrier lifetime is comparable to the THz field pulse or even longer, the measured photocurrent will no longer be directly proportional to the electric field of the THz pulse and further mathematical processing is required to get the actual pulse shape. The reconstruction technique to get the actual THz pulse shape $E(\tau)$ from the recorded current $I(\tau)$ with a slow carrier decay photoconductor is explained in reference [9]. There have been attempts to improve the THz detection ability of SI-GaAs as a photoconductor by implanting it with different ions, like Ar, N. [7, 10]. However, in most cases these implantations lead to doping of the material and then, due to the dark conductivity of the residual carriers present in the material, the THz detection efficiency is reduced.

When using photoconductors for THz detection, it is important to know the carrier recombination lifetime $\tau_{rec}$ of the photoconductor as the recorded pulse shape depends on $\tau_{rec}$. To estimate the carrier lifetime $\tau_{rec}$, optical pump-probe reflection measurements were performed and the results are shown in fig 2. The carrier lifetime of non-irradiated SI-GaAs was found to be $\tau_{rec} \sim 70$ ps, whereas that of the irradiated samples were 6 ps, 2.2 ps, 1.8 ps and 0.55 ps for estimated dosages $10^{12}$, $10^{13}$, $10^{14}$ and $10^{15}$ cm$^{-2}$, respectively. Finally the photoconducting THz detectors were tested in a standard THz- time domain spectroscopy (TDS) set up. A commercial photoconductive antenna source (Batop-iPCA) made on LT-GaAs was used as THz source. The generated THz pulse was focused on these detectors using off-axis parabolic mirrors and a small fraction (~5 mW) of the 300 mW, 10fs, 800 nm laser pulse was also focused on these detectors after passing through a delay station. The photocurrents recorded as a function of delay times are plotted in fig. 3 for different detectors. The THz pulse recorded with the electro optic technique, using a ~1 mm thick <110> ZnTe crystal, is also shown (as the black curve) in fig. 3(f). To a good approximation, this signal can be considered as the actual electric field profile $E(t)$ of the THz pulse. Theoretical photocurrent curves $I(\tau;\tau_{rec})$ for a given recombination lifetime can then be obtained by a convolution of this $E(t)$ with the carrier density response function

$$n(t-\tau) = n_0 \, exp[-(t-\tau)/\tau_{rec}], \qquad (2)$$

according to Eq.(1). In fig.s 3(a-e) the recorded experimental $I(\tau)$ from the photodetector with different irradiation doses (full lines) are plotted together with the corresponding theoretical curves $I(\tau;\tau_{rec})$ using the corresponding carrier lifetimes $\tau_{rec}$ obtained from the pump-probe reflection measurements (dashed lines). The agreement between measured and theoretical curves is very good. As expected, the photocurrent pulse shape recorded from non-irradiated SI-GaAs ($\tau \sim$ 70 ps) is way different from the actual THz pulse shape recorded from ZnTe. The noise level is also high because of the long lived charge carriers. As we increase the irradiation dose, the decreased carrier life time improves the recorded photo current pulse shape (fig. 3(a-e); from fig. 3(f) the increasingly faithful shape is best visible). Also, the signal to noise ratio increases from ~2-3 to ~ 50 for the maximum dose. To get the frequency dependent response of these detectors, fast Fourier transforms (FFTs) of the recorded pulses are plotted in fig. 4(a-c). The noise levels are indicated by horizontal dashed lines and the S/N for different frequencies can be easily observed.

In summary, carbon irradiation of SI-GaAs improves its properties as a material for THz generation and detection in two ways. First, it increases the thermal breakdown voltage of THz *sources* fabricated from irradiated SI-GaAs. Thus, it has enabled us to apply higher bias voltages (~ 150 V, in comparison to ~ 35-40 V) on our THz sources , resulting in an increase of the emitted THz amplitude by a factor ~ 12, as reported in Ref.[8]. In fact, the efficiency of these THz sources can match the performance of LT-GaAs emitters. Second, carbon irradiation results in an improvement of the *detection* of THz pulses from photoconductive sources made on SI-GaAs. The irradiated SI-GaAs detectors exhibit a strongly enhanced S/N ratio. Also, they are able to detect higher THz frequency components in comparison to non-irradiated SI-GaAs detectors (see table 1.).

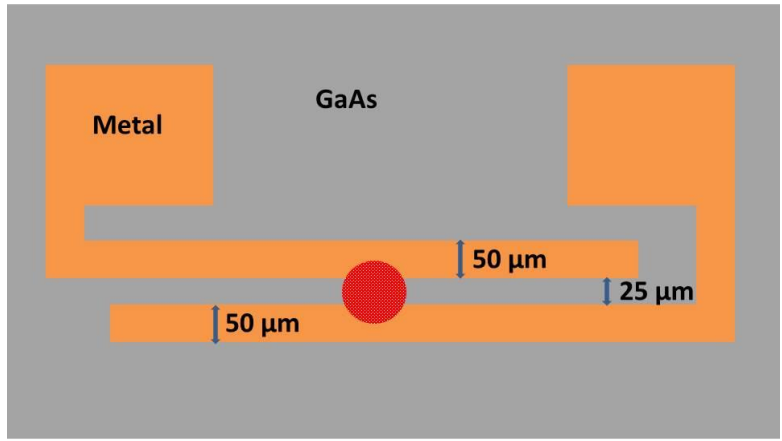

**Figure 1.** Schematic diagram of the electrodes on a photoconductor used for THz pulse detection. The THz pulse to be detected will be focused on the same spot of optical carrier excitation with almost matching arrival time of the 800 nm pulse. The photogenerated charge carriers will move in the presence of the electric field of the THz pulse and a current flow will be measured at the

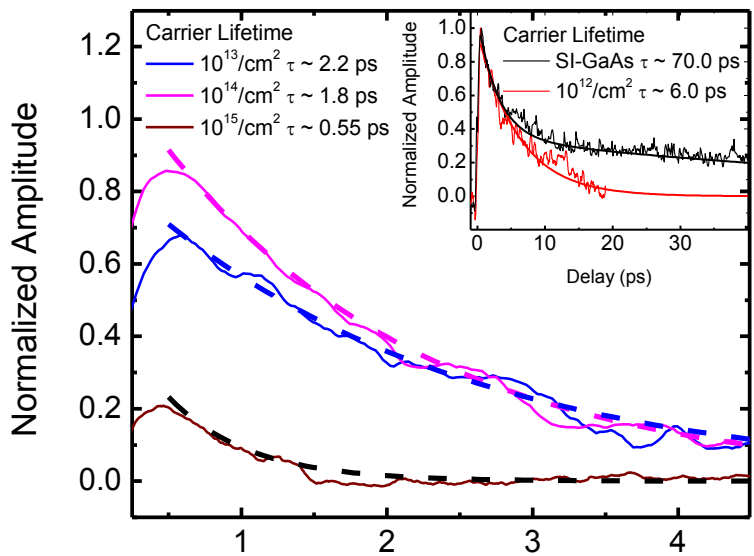

**Figure 2.** Optical NIR time resolved pump-probe reflection curves of non-irradiated and irradiated SI-GaAs. The dash-lines are fits to a single exponential decay. The inset shows the non-irradiated sample with mostly a slow fall (~70 ps), except for the beginning of the signal, while the low irradiated sample exhibits already a comparatively rapid decay (~6 ps).

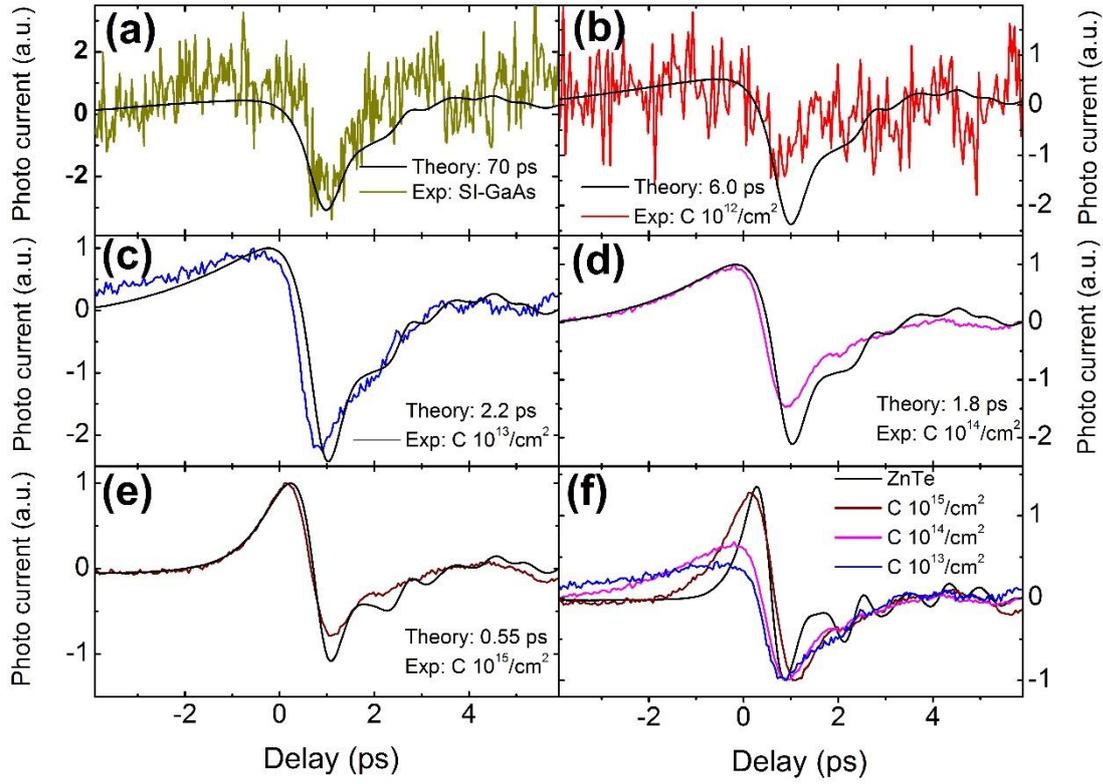

**Figure 3.** THz pulses recorded from different carbon irradiation dose photoconductive antennas (PCAs) with their theoretically predicted pulse shapes (a-e). In (f) the THz pulse shape recorded from the 3 highest dosages PCA detector is plotted together with the pulse shape recorded from the ZnTe electro optic crystal (continuous black curve).

**Table 1.**

| Irradiation Dose (ions/cm$^2$) | Estimated Carrier lifetime (ps) | S/N for Detection | Detecion Bandwidth (THz) |
|---|---|---|---|
| 0 | ~ 70 | ~ 2-3 | < 0.5 |
| ~ $10^{12}$ | ~ 6 | ~ 2-3 | < 0.5 |
| ~ $10^{13}$ | ~ 2.2 | ~ 20 | < 1 |
| ~ $10^{14}$ | ~ 1.8 | ~ 40 | < 1 |
| ~ $10^{15}$ | ~ 0.55 | ~ 50 | < 1.5 |

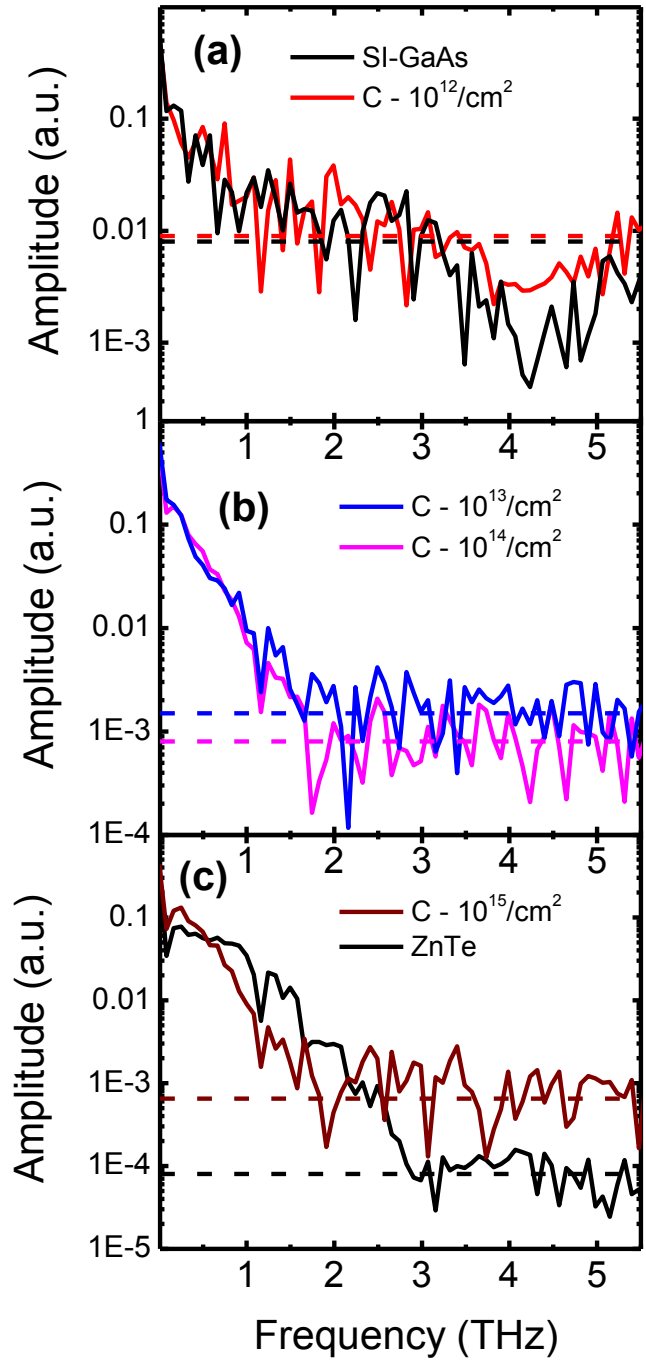

**Figure 4.** FFTs of the THz pulses recorded from different carbon irradiation dose photoconductive antennas. The dotted lines are indicating the noise level.